\documentclass[preprintnumbers, twocolumn,showkeys,superscriptaddress,showpacs,
nofootinbib,prd,notitlepage]
{revtex4-1}
\usepackage{bbm,pifont}
\usepackage{amssymb}
\usepackage{amsmath}
\usepackage{physics}
\usepackage{enumitem}
\usepackage{amsfonts}
\usepackage{epsfig}
\usepackage{graphicx}
\usepackage{tabularx}
\usepackage{color}
\usepackage[dvipsnames]{xcolor}
\usepackage{slashed}
\usepackage{multirow}
\usepackage[colorlinks=true, linkcolor=blue, citecolor=blue, urlcolor=blue]{hyperref}

\begin{document}

\title{Quantum Brownian Motion as a Classical Stochastic Process in Phase Space}

\author{Dmitriy Kondaurov}
\email[]{kd8466@mail.ru}
\affiliation{Russian Quantum Center, 30 Bolshoy Boulevard, building 1, Skolkovo Innovation Center territory, Moscow, 121205, Russia} 
\affiliation{Moscow Institute of Physics and Technology, Institutsky lane 9, Dolgoprudny, Moscow region, 141700, Russia}
\author{Evgeny Polyakov}
\email[]{evgenii.poliakoff@gmail.com}
\affiliation{Russian Quantum Center, 30 Bolshoy Boulevard, building 1, Skolkovo Innovation Center territory, Moscow, 121205, Russia}

\begin{abstract}
We establish that the exact quantum dynamics of a Brownian particle in the Caldeira-Leggett model, with at most quadratic external potential, can be mapped, at any temperature, onto a classical, non-Markovian stochastic process in phase space. Starting from a correlated thermal equilibrium state between the particle and bath, we demonstrate that this correspondence is exact for quadratic potentials under arbitrary quantum state preparations of the particle itself. Our approach allows to consider arbitrary initial quantum states—including highly non-classical superpositions—which are incorporated via their Wigner functions, which serve as statistical weights for trajectory ensembles. Furthermore, the formalism naturally accommodates external manipulations and measurements modeled by preparation functions acting at arbitrary times, enabling the simulation of complex driven-dissipative quantum protocols. For more general, smooth potentials, we identify a natural small parameter: the density matrix becomes strongly quasidiagonal in the coordinate representation, with its off-diagonal width shrinking as the bath's spectral cutoff increases, suggesting a controlled parameter for a possible approximation.
\end{abstract}
\maketitle
\section{Introduction}

Brownian motion—the erratic dynamics of a microscopic system interacting with a vast environment—is a cornerstone of statistical physics. Its classical description in terms of a Langevin equation\cite{Langevin1908}, combining friction and random noise, is ubiquitous across chemical kinetics, biophysics, and materials science \cite{Hanggi1990}. On smaller scales where quantum effects become significant, such as in low-temperature defect dynamics, quantum diffusion or tunneling \cite{Leggett1984QuantumTunneling, BEC, torres2022surface, gottwald2015applicability, Caldeira1983, Diósi2020}, a fundamental question arises: how can Brownian motion be consistently quantized? This problem, known as quantum Brownian motion (QBM), is notoriously difficult. The principles of quantum mechanics forbid a naive quantization of classical dissipative equations, leaving a microscopic description of the system coupled to a macroscopic environment as the only rigorous path. However, the exponential growth of the Hilbert space dimension makes direct simulation intractable, creating a pressing need for a reduced description in terms of the Brownian particle alone.

The first microscopically consistent model for QBM was introduced by Caldeira and Leggett  \cite{Caldeira1983path}, representing the environment as a bath of independent harmonic oscillators linearly coupled to the particle. A foundational and well-known result of this model is that the Heisenberg equations of motion for the particle operators are formally identical to the classical generalized Langevin equation \cite{Weiss_2012, gardiner2004quantum}. This formal correspondence suggests that the quantum dynamics might be representable in classical terms, yet translating this operator identity into a practical, all-temperature description of the quantum state has remained a challenge.

Early attempts to derive reduced equations of motion often resorted to approximations valid only in limiting regimes. For instance, master equations of Lindblad form \cite{Lindblad1976}, which are common in quantum optics, can be derived for QBM but are strictly valid only in the high-temperature limit ($T \gg \hbar \Lambda$, where $\Lambda$ is a  characteristic scale of spectrum width; we set $k_B = 1$ throughout) \cite{breuer, Diosi1993}. At low temperatures, these equations fail to capture essential quantum features such as zero-point fluctuations and can even violate physical principles \cite{Vacchini2000, Vacchini2001}. Despite this limitation, their relative simplicity has sustained their use in the literature. The more general, Redfield equation also has the same problem with complete positivity \cite{Tanimura2006, Tanimura2020}. For Caldeira-Leggett model with no more than quadratic potential it is possible to write exact masterequation \cite{Hu1992, Intravaia2003}, but only for a factorized initial state, and it does not permit quantum operations during the evolution. The exact path-integral solutions \cite{Feynman1965, Feynman1963} for the Caldeira-Leggett model, while powerful \cite{Grabert1988, Schramm1986, Hu1992, Kamenev2023}, are analytically cumbersome, especially for arbitrary preparations and in the presence of interventions.

Among exact methods, the stochastic Schr\"odinger equation \cite{Strunz1999b, Strunz1999, Diósi1997, Diosi1998, Suess2014} provides a general and highly efficient framework. It explicitly constructs a stochastic process using noise with a nonlocal time dependence. While such a construction is always possible, it becomes computationally demanding for unconfined potentials, where the dimension of the required state space grows large.

The hierarchical equations of motion (HEOM) offer another numerically exact solution for the reduced density matrix with very broad applicability \cite{Tanimura2020}. Here too, the numerical cost grows with the state-space dimension when many levels are involved. An additional and well-known difficulty arises at low temperatures, where the slow algebraic decay of bath correlations significantly increases computational demands, even with recent improvements in correlation decomposition \cite{Tanimura2020, Cui2020, Bai2024}.

These methods are powerful and well-established. For the specific case of the Caldeira-Leggett model, however, they face specific obstacles rooted in its continuous-variable nature. Even in its simplest, quadratic form, the Caldeira-Leggett model thus presents a tension: the underlying Hamiltonian is integrable, yet its quantum simulation at arbitrary temperature remains nontrivial. Resolving this tension by finding a simple computational framework for the quadratic case is a natural first step, and may later guide approximations beyond it. This motivates the development of a more specialised computational approach that leverages the intrinsic structure of the Caldeira-Leggett model to achieve greater numerical efficiency in this particular setting.

In this work, we demonstrate that the exact quantum dynamics of the Caldeira-Leggett model with at most quadratic external potential admits a complete and practical representation as a classical, non-Markovian stochastic process in phase space, valid at all temperatures. We establish that, starting from a correlated thermal equilibrium state, the reduced quantum dynamics for a particle in a quadratic potential is identical to an ensemble of trajectories governed by a classical generalized Langevin equation. The quantum nature of the environment is encoded exactly in the statistics of a colored noise that obeys the quantum fluctuation-dissipation theorem. For smooth, non-quadratic potentials, we identify a natural small parameter—the off-diagonal width (coherence length) of the equilibrium reduced density matrix—which shrinks with increasing bath spectral cutoff. We expect that it controls the accuracy of the approximation.

The primary value of this mapping is its profound simplification: it reduces the formidable problem of simulating an open quantum system with infinitely many degrees of freedom to the tractable task of generating and averaging classical stochastic trajectories. Our associated numerical technique accommodates arbitrary initial quantum states—including non-classical superpositions—via their Wigner representations, and can incorporate external manipulations through preparation functions. Crucially, the numerical complexity of our approach is essentially independent of temperature, allowing it to operate seamlessly from the classical high-temperature regime down to the deeply quantum, zero-temperature limit without increasing numerical complexity. This provides a unified computational tool that is conceptually straightforward and offers a versatile alternative to more complex contemporary methods for this specific class of problems.

The remainder of this paper is structured as follows: In Sec.~\ref{sec:Em-Dec_His}, we review the Caldeira-Leggett model. In Sec.~III, we present the core result—the exact classical-stochastic mapping—and detail its proof. Section IV describes the numerical implementation. In Sec. V, we validate the method against known analytical results and contrast it with high-temperature master equations, highlighting the crucial all-temperature behavior. Section VI discusses the extension to non-quadratic potentials via the identified small parameter. We conclude in Sec.~\ref{sec:concl}.

\section{\label{sec:Em-Dec_His}Microscopic model of quantum Brownian motion}

\subsection{Caldeira-Leggett Model}

The paradigmatic model for quantum Brownian motion is the Caldeira-Leggett model \cite{Caldeira1983}, which represents a one-dimensional Brownian particle of mass \(m\) bilinearly coupled to a macroscopic environment. The environment is modeled as a bath of independent harmonic oscillators. The Hamiltonian for the total system is
\begin{equation}
\begin{gathered}
    H = \frac{p^2}{2m} + V(x) + \sum_{i} \left[\frac{p_i^2}{2m_i} + \frac{1}{2} m_i \omega_i^2 x_i^2\right] \\- x \sum_{i} c_i x_i + x^2 \sum_{i} \frac{c_i^2}{2 m_i \omega_i^2},
\end{gathered}
\end{equation}
where \(x, p\) are the coordinate and momentum of the particle, and \(x_i, p_i, m_i, \omega_i\) are the coordinates, momenta, masses, and frequencies of the bath oscillators. The constants \(c_i\) quantify the coupling strength. The first two terms constitute the Hamiltonian of the isolated particle. The third term is the Hamiltonian of the isolated bath. The fourth term describes the linear system-bath interaction, and the final counter-term is necessary to ensure that there are no additional forces acting on the free particle at rest \cite{Caldeira1983, Weiss_2012}.

\subsection{Joint State and Preparation}
The total system's quantum state is described by a density matrix \(\rho_{SB}\). We consider the physically relevant scenario where, at an initial time \(t=0\), the total system is prepared starting from a joint thermal equilibrium state at inverse temperature \(\beta = 1/T\):
\begin{equation}
    \rho_{SB}^\beta = Z^{-1} e^{-\beta H}, \label{eq:joint_thermal}
\end{equation}
where \(H\) is the full Hamiltonian in Eq.~(1) and \(Z\) is the partition function. This state contains all system-bath correlations consistent with equilibrium.
An arbitrary preparation of the Brownian particle's state is then enacted via an operation acting solely on its Hilbert space \cite{Grabert1988}. This is described by a set of preparation operators \(\{O_j\}\), yielding the initial state
\begin{equation}
    \rho_{SB}(0) = \sum_j O_j \, \rho_{SB}^\beta \, O_j^\dagger. \label{eq:prepared_state}
\end{equation}

In coordinate representation it is described by preparation function
\begin{equation}
    \lambda(x, x'|\bar x, \bar x') = \sum_{j} \langle x' | O_j | \bar x' \rangle \langle \bar x | O_j^\dag | x \rangle.
\end{equation}

This formalism can be generalized to describe interventions (e.g., measurements or controlled manipulations) at intermediate times \(t_k\).

\subsection{Reduced Quantum Dynamics with Intermediate Interventions}
The formalism of preparation operators naturally extends to describe a sequence of interventions---such as measurements, unitary kicks, or state preparations---applied to the particle at specific intermediate times. Consider a sequence of times \( 0 < t_1 < t_2 < \dots < t_N \). At each time \( t_k \), an intervention is described by a set of operators \(\{ O_j^{(k)} \}\) acting solely on the particle's Hilbert space. The quantum state evolves as follows:
\begin{enumerate}
    \item {Initial Preparation:} Starting from the joint thermal state \( \rho_{SB}^\beta \) from Eq.~\eqref{eq:joint_thermal}, an initial preparation is performed at \( t=0 \):
    \begin{equation}
        \rho_{SB}(0^{+}) = \sum_{j_0} O_{j_0}^{(0)} \, \rho_{SB}^\beta \, \big(O_{j_0}^{(0)}\big)^\dagger.
    \end{equation}
    \item {Unitary Evolution:} The state evolves unitarily under the full Hamiltonian until the first intervention:
    \begin{equation}
        \rho_{SB}(t_1^{-}) = U(t_1) \, \rho_{SB}(0^{+}) \, U^\dagger(t_1), \quad \text{where } U(t) = e^{-i H t / \hbar}.
    \end{equation}
    \item {Intervention at \( t_k \):} At time \( t_k \), the intervention operators are applied:
    \begin{equation}
        \rho_{SB}(t_k^{+}) = \sum_{j_k} O_{j_k}^{(k)} \, \rho_{SB}(t_k^{-}) \, \big(O_{j_k}^{(k)}\big)^\dagger.
    \end{equation}
    The state then continues its unitary evolution until the next intervention or the final time.
    \item {Final State and Observables:} After the last intervention at \( t_N \), the state evolves to the any time $t > t_N$:
    \begin{equation}
        \rho_{SB}(t) = U(t - t_N) \, \rho_{SB}(t_N^{+}) \, U^\dagger(t - t_N).
    \end{equation}
    The physically relevant object is the reduced density matrix of the particle at \( t \), obtained by tracing out the bath:
    \begin{equation}
        \rho_S(t) = \Tr_B \big[ \rho_{SB}(t) \big].
    \end{equation}
    The expectation value of any particle observable \( \hat{A} \) at an arbitrary time \( t \) within the interval \( [0, t] \) is defined by the state at that time. The observable \( \hat{A} \) acts solely on the particle's Hilbert space. The time \( t \) may precede, follow, or lie between the intervention times \( \{t_k\} \). The general expression is
\begin{equation}
    \langle \hat{A}(t) \rangle = \Tr_{SB}\big[ \rho_{SB}(t) \, (\hat{A} \otimes \hat{I}_B) \big] = \Tr_S\big[ \rho_S(t) \, \hat{A} \big],
\end{equation}
where \( \rho_{SB}(t) \) is the total density matrix at time \( t \), constructed according to the sequence of unitary evolution and interventions described above, \( \hat{I}_B \) is the identity operator on the bath Hilbert space, and \( \rho_S(t) = \Tr_B[ \rho_{SB}(t) ] \) is the reduced density matrix of the particle.
\end{enumerate}
This framework provides a complete prescription for computing quantum observables under arbitrary sequences of preparations and interventions within the Caldeira-Leggett model. The key result of our work is that this complex, non-Markovian quantum evolution can be mapped exactly onto a classical stochastic process for quadratic potentials, as detailed in the following sections.

\subsection{Reduced Equation of Motion}
\label{sec:red_eq}
The Heisenberg equations of motion are
\begin{gather}
    m \ddot{x}(t) = -V'(x(t)) + \sum_i c_i x_i(t) - x(t) \sum_i \frac{c_i^2}{m_i \omega_i^2}, \label{eq:particle_eq} \\
    m_i \ddot{x}_i(t) + m_i \omega_i^2 x_i(t) 
    = c_i x(t). 
    \label{eq:bath_eq}
\end{gather}

Equation \eqref{eq:bath_eq} for the bath oscillators is a linear inhomogeneous equation. Its formal solution, substituting the particle's trajectory \(x(t)\), is
\begin{equation}
\begin{aligned}
    x_i(t) = x_i(0) \cos(\omega_i t) + \frac{p_i(0)}{m_i \omega_i} \sin(\omega_i t) \\+ \frac{c_i}{m_i \omega_i} \int_0^t \sin[\omega_i (t-\tau)] x(\tau) \, d\tau. \label{eq:bath_solution}
\end{aligned}
\end{equation}
Substituting Eq.~\eqref{eq:bath_solution} into Eq.~\eqref{eq:particle_eq} yields a closed, non-Markovian equation for the particle alone:
\begin{equation}
    m \ddot{x}(t) = -V'(x(t)) - \int_0^t M(t-\tau) \dot{x}(\tau) \, d\tau - x(0) M(t) + \xi(t). \label{eq:gen_langevin}
\end{equation}
This is a generalized Langevin equation (GLE). Its constituents are:
\begin{itemize}
    \item The memory kernel (friction kernel):
    \begin{equation}
        M(t) = \sum_i \frac{c_i^2}{m_i \omega_i^2} \cos(\omega_i t). \label{eq:memory_discrete}
    \end{equation}
    \item The stochastic force (noise):
    \begin{equation}
        \xi(t) = \sum_i c_i \left[ x_i(0) \cos(\omega_i t) + \frac{p_i(0)}{m_i \omega_i} \sin(\omega_i t) \right]. \label{eq:noise_discrete}
    \end{equation}
\end{itemize}
The term \(-x(0)M(t)\) arises from the non-translational-invariant \cite{PhysRev2} preparation of the initial bath state relative to the particle. It is non-physical for describing steady-state Brownian motion and is typically removed by assuming the bath was in equilibrium with the particle held at position \(x(0)\) for all past times. This is equivalent to redefining the initial bath coordinates as \(x_i(0) \to x_i(0) + \frac{c_i}{m_i \omega_i^2} x(0)\), which cancels the term exactly. We adopt this physically consistent preparation, yielding the standard GLE:
\begin{equation}
    m \ddot{x}(t) = -V'(x(t)) - \int_0^t M(t-\tau) \dot{x}(\tau) \, d\tau + \xi(t). \label{eq:gle_clean}
\end{equation}

% \subsection{Spectral Density and Thermodynamic Limit}
In the thermodynamic limit of a continuous bath, the system properties are encoded in the spectral density
\begin{equation}
    J(\omega) = \frac{\pi}{2} \sum_i \frac{c_i^2}{m_i \omega_i} \delta(\omega - \omega_i). \label{eq:spectral_density_def}
\end{equation}
The memory kernel is then expressed as
\begin{equation}
    M(t) = \frac{2}{\pi} \int_0^{\infty} d\omega \, \frac{J(\omega)}{\omega} \cos(\omega t). \label{eq:memory_cont}
\end{equation}
For an ohmic bath with an exponential cutoff on $\Lambda$ (corresponding to a time scale $\varepsilon = \Lambda^{-1}$), \(J(\omega) = \gamma m \omega e^{-\varepsilon \omega}\), where \(\gamma\) is the friction coefficient. The corresponding memory kernel is \(M(t) = (2 m \gamma / \pi) \varepsilon / (\varepsilon^2 + t^2)\). In the Markovian limit  $\varepsilon \ll \min \{\gamma^{-1}, \tau_S \}$, where $\tau_S$ is characteristic time scale of the particle motion (e.g. $\Omega^{-1}$ for the oscillator with frequency $\Omega$), \(M(t) \to 2 m \gamma \delta(t)\), and Eq.~\eqref{eq:gle_clean} reduces to the standard Langevin equation:
\begin{equation}
    m \ddot{x}(t) = -V'(x(t)) - m \gamma \dot{x}(t) + \xi(t).
\end{equation}

\subsection{Fluctuation-Dissipation Theorem}
Assuming bath in thermal equilibrium \(\rho_B \propto \exp(-H_B / T)\), the noise is Gaussian with zero mean and its autocorrelation function is given by the quantum fluctuation-dissipation theorem (FDT) \cite{breuer}:

\begin{equation}
    \langle \xi(t)  \xi(s) \rangle = \frac{1}{2} \big( \langle \{\xi(t), \xi(s) \} \rangle + \langle [\xi(t), \xi(s) ] \rangle \big)
\end{equation}

where

\begin{multline}
\label{eq:FDT}
    \langle \{\xi(t), \xi(s) \} \rangle = \frac{2 \hbar}{\pi} \int_{0}^{\infty} d\omega \, J(\omega) \coth \left( \frac{\hbar \omega}{2 T} \right) \cos [\omega (t-s)] ;\\
     \langle [\xi(t), \xi(s) ] \rangle =[\xi(t), \xi(s) ] =-i \frac{2 \hbar}{\pi} \int_{0}^{\infty} d\omega \, J(\omega)  \sin [\omega (t-s)]
% \tag{27}
\end{multline}

At \(T=0\), \(\coth(\hbar \omega / (2 T)) \to 1\), leading to zero-point fluctuation noise. For an ohmic bath \(J(\omega) = \gamma m \omega e^{-\varepsilon \omega }\), this yields the non-Markovian correlation function
\begin{equation}
    \langle \xi(t) \xi(0) \rangle_{T=0} = \frac{m \gamma \hbar}{\pi} \frac{\varepsilon^2 - t^2}{(\varepsilon^2 + t^2)^2},
\end{equation}
which is clearly distinct from the classical white-noise limit \cite{gardiner2004quantum}.

\subsection{Regimes of Caldeira-Leggett model}
\label{regimes}

Brownian particle in the frame of Caldeira-Leggett model has different features of behavior which depend on temperature and spectral density\cite{Grabert1988}. Consider spectral density in the following form:
\begin{equation}
    J(\omega) = g_{\alpha} \omega^{\alpha} \exp{-\frac{\omega}{\Lambda}}.
\end{equation}

Those spectral densities can be divided into the following regimes. $\alpha = 1$ corresponds to ohmic case, which was considered above. $0<\alpha < 1$ (subohmic), $1<\alpha \leq 2$ and $\alpha > 2$.

To proceed we will need to understand the character of asymptotic for non-equilibrium states in the different cases, particularly when an arbitrary initial state approaches equilibrium with $t \xrightarrow[]{}\infty$.

Thus, for $\alpha<1$ and $T=0$ we have localization in coordinate space, for $\alpha = 1$ and $T=0$ - logarithmic (ultraslow) diffusion,  for $\alpha > 2$, asymptotically, the Brownian particle moves freely with a nonequilibrium distribution, depending on its initial state. Note that true equilibrium distribution has factorized form (in Wigner representation) $W(x, p, \infty) = w_\infty(x) w_\infty(p)$. In case with $\alpha > 2$ distribution on infinite times does not look like this.

\section{Wigner Picture and the Classical-Quantum Correspondence}

 Consider external potential $V(x)$ which is no more than quadratic. Thus the full Hamiltonian of the Caldeira-Leggett model is quadratic in both the particle and bath coordinates. This fundamental property leads to a profound simplification in the phase-space description of the quantum dynamics: the quantum Liouville equation for the Wigner function of the total system is formally identical to the classical Liouville equation for the joint probability distribution in phase space \cite{Weiss_2012}. 

Wigner function is defined by
\begin{equation}
    W(r, p, t) = \int dq e^{-\frac{i}{\hbar} p q} \rho (r, q, t),
\end{equation}
where we introduced ''center mass'' variables $r = \frac{x + x'}{2}, \, \, q = x - x'$ for density matrix $\rho(x,x',t)$ in coordinate representation.

\subsection{From Wigner Dynamics to Stochastic Trajectories}

Let \(W^{SB}(r, p, \{ r_i, p_i \}, t)\) denote the Wigner function of the total system (particle + bath). $(r, p)$ - particle, $(r_i, p_i)$ - all oscillators. For a quadratic Hamiltonian, its evolution is given by
\begin{equation}
    \frac{\partial W^{SB}}{\partial t} = \{ H, W^{SB} \}_{\text{P.B.}},
    \label{eq:liouville_wigner}
\end{equation}
where \(\{ \cdot, \cdot \}_{\text{P.B.}}\) is the classical Poisson bracket. Equation \eqref{eq:liouville_wigner} is the classical Liouville equation. It means that each point in phase space evolves according to the classical equations of motion \cite{Moyal1949}, so the dynamics of full system can be represented by an ensemble of deterministic trajectories in the total phase space, evolving under Hamilton's equations derived from \(H\), which are exactly coincide with equations (\ref{eq:particle_eq}, \ref{eq:bath_eq}), but for classical functions $x(t)$, $x_i(t)$. We also demonstrate it in Appendix \ref{pathint} using path integral approach.

This establishes a rigorous trajectory-based interpretation of the exact quantum dynamics for the Caldeira-Leggett model. 
\subsection{The Reduced Stochastic Process for the Brownian Particle}

The Hamiltonian structure of the Caldeira-Leggett model allows the bath degrees of freedom to be eliminated analytically from the joint trajectories, exactly as in the section \ref{sec:red_eq}. We start from the equilibrium state of the full system, so if we want to work only with trajectories we can expand time interval into the past, using the property $[H^{SB}, \rho_\beta^{SB}] = 0$ of the equilibrium state. Thus, for negative times we will work with equilibrium trajectories, corresponding to a given realization of noise. For each sampled trajectory, the motion of the Brownian particle obeys the generalized Langevin equation (GLE):
\begin{equation}
    m \ddot{r}(t) = -V'(r(t)) - \int_{-T}^{t} M(t-\tau) \dot{r}(\tau)  d\tau + \xi_W(t),
    \label{eq:gle_trajectory}
\end{equation}
where the memory kernel \(M(t)\) is defined in Eq.~\eqref{eq:memory_cont}. The crucial difference from the classical case lies in the statistics of the stochastic force \(\xi(t)\).

The force \(\xi(t)\) is constructed from the initial conditions of the bath oscillators \(\{r_i(-T), p_i(-T)\}\):

\begin{equation}
\begin{aligned}
\xi(t) = \sum_i c_i \Big[ & r_i(-T) \cos(\omega_i (t+T)) \\
& + \frac{p_i(-T)}{m_i \omega_i} \sin(\omega_i (t+T)) \Big].
\end{aligned}
\end{equation}

We suppose the equilibrium state for full system $\rho^{SB}$ with corresponding $W^{SB}$. Because these initial conditions are drawn from the quantum thermal Wigner function, the noise correlation is given by the quantum fluctuation-dissipation theorem for noise operator (\ref{eq:FDT}).

The symmetric (real) part of this correlation, which governs observable averages, for $\xi_W$ is
\begin{multline}
    \langle  \xi_{W}(t) \xi_{W}(s)  \rangle = \frac{1}{2} \langle \{ \hat \xi(t), \hat \xi(s) \} \rangle =\\ = \frac{1}{\pi} \int_0^{\infty} d\omega \, J(\omega) \hbar \coth\left(\frac{\hbar \omega}{2 T}\right)  \cos[\omega (t-s)].
    \label{eq:noise_quantum_symmetric}
\end{multline}

Thus, the exact quantum dynamics of the particle is reproduced by an ensemble of trajectories solving the classical GLE \eqref{eq:gle_trajectory}, where the only ''quantum'' ingredient is the noise statistics derived from the initial quantum Wigner distribution of the bath.

\subsection{Incorporating State Preparation and Interventions}

The preparation formalism of Section 2.2 is incorporated naturally within this trajectory picture. A preparation or intervention at time \(t_k\), described by operators \(\{O_\alpha^{(k)}\}\), corresponds to a non-classical update of the particle's Wigner function. In our stochastic framework, this is implemented by assigning a weight \(w_\alpha^{(j)}\) to each trajectory \(j\) at time \(t_k\).

The weight is derived from the preparation function \(\lambda\) in the Wigner representation \cite{Hanggi1993}. For a trajectory that has phase-space coordinates \((\bar{r}, \bar{p})\) just before \(t_k\) and \((r_0, p_0)\) just after, the weight contributed by the intervention is
\begin{equation}
    w^{(j)}(t_k) = \lambda(r_0^{(j)}, p_0^{(j)} | \bar{r}^{(j)}, \bar{p}^{(j)}),
\end{equation}
where $\lambda(r_0, p_0 | \bar{r}, \bar{p}) = \int d\bar q dq_0 e^{\frac{i}{\hbar}(\bar q \bar p - q_0 p_0)} \lambda (r_0, q_0 | \bar r, \bar q) $. The point $(r_0, p_0)$  is assumed to be uniformly distributed in phase space, and the conditional probability of a trajectory break between points $(\bar r, \bar p)$ and $(r_0, p_0)$ is taken into account in the weight. This allows for working with essentially quantum preparation functions that have negative regions and are not interpreted as probabilities.

For a sequence of interventions, the total weight for a trajectory is the product of weights from all events. The expectation value of an observable at time \(t\) is then computed by a weighted average over the ensemble:
\begin{equation}
    \langle \hat{O}(t) \rangle = \frac{ \sum_{j=1}^{N} \left( \prod_{\{t_k < t\}} w^{(j)}(t_k) \right) O_W\left(r^{(j)}(t), p_S^{(j)}(t)\right) }{\sum_{j=1}^{N} \left( \prod_{\{t_k < t\}} w^{(j)}(t_k) \right) }.
    \label{eq:observable_weighted_average}
\end{equation}
This formulation provides a complete and practical method for simulating the non-Markovian quantum dynamics of the Brownian particle under arbitrary sequences of preparations and measurements, provided the potential is quadratic. The numerical implementation of this stochastic approach is detailed in the next section.

\section{Numerical Method: Stochastic Monte Carlo Simulation}

The theoretical correspondence established in the previous sections---mapping the quantum dynamics of the Caldeira-Leggett model  with at most a quadratic Hamiltonian to a classical stochastic process---lends itself directly to numerical implementation via a Monte Carlo technique. The core algorithm involves: (i) sampling initial  equilibrium trajectories, (ii) generating stochastic trajectories by solving the generalized Langevin equation (GLE), and (iii) computing quantum observables via weighted ensemble averages that account for state preparations.

\subsection{Generation of Quantum Noise}
The key ingredient for simulating the quantum GLE is the generation of stochastic force trajectories \(\xi^{(j)}(t)\) whose two-time correlation satisfies the quantum fluctuation-dissipation theorem:
\begin{equation}
    \langle \xi(t) \xi(s) \rangle = \frac{1}{\pi} \int_0^{\infty} d\omega \, J(\omega) \hbar\coth\left(\frac{\hbar \omega}{2 T}\right) \cos[\omega (t-s)].
    \label{eq:noise_target}
\end{equation}
A numerically efficient method is to synthesize \(\xi(t)\) in the frequency domain. We express the noise as
\begin{equation}
    \xi(t) = \sqrt{\frac{1}{2\pi}} \int_{-\infty}^{\infty} d\omega \, \sqrt{N(\omega)} \, z(\omega) e^{-i\omega t},
    \label{eq:noise_spectral}
\end{equation}

where \(N(\omega) = \hbar J(|\omega|)\coth(\hbar |\omega| / (2 T))\) and \(z(\omega)\) is a complex Gaussian auxiliary noise with statistics
\begin{multline}
    z^*(\omega) = z(-\omega), \quad \langle z(\omega) z^*(\omega') \rangle = \delta(\omega - \omega'), \\ \langle z(\omega) z(\omega') \rangle = 0.
    \label{eq:z_statistics}
\end{multline}
The discretized numerical implementation uses a finite frequency grid \(\omega_k = k \Delta \omega\) for \(k = -N, \dots, N\). For each \(k\), we generate independent complex random variables \(\eta_k\) and \(\zeta_k\) from a standard normal distribution \(\mathcal{N}(0,1)\) and construct
\begin{equation}
    z_k = \frac{1}{\sqrt{2}} (\eta_k + i \zeta_k), \quad \text{for } k > 0,
\end{equation}
with \(z_0\) being real (\(\eta_0\)) and \(z_{-k} = z_k^*\) to satisfy Eq.~\eqref{eq:z_statistics}. The noise time series at discrete times \(t_n\) is then computed via the inverse discrete Fourier transform:
\begin{equation}
    \xi(t_n) = \sqrt{\frac{\hbar \Delta \omega}{2\pi}} \sum_{k=-N}^{N} \sqrt{N(\omega_k)} \, z_k \, e^{-i \omega_k t_n}.
    \label{eq:noise_discrete}
\end{equation}
This procedure generates a stationary Gaussian process whose ensemble average reproduces Eq.~\eqref{eq:noise_target} with accuracy controlled by the spectral resolution \(\Delta \omega\) and cutoff \(N \Delta \omega\).

Let us estimate the error caused by the frequency discretization $\Delta \omega$ and the finite cutoff $\omega_m = N\Delta\omega$ (in the case of exponentially decaying density, there is nowhere zero, so we must introduce additional cutoff $\omega_m \gg \varepsilon^{-1}$). Consider integral (\ref{eq:noise_target}). For example, for the ohmic spectral density, this integral will have a relative error of the order of $\varepsilon \omega_m e^{-\varepsilon \omega_m}$. The discretization error is mainly defined by error of rectangle method of numerical integration, which is proportional to $\omega_m \Delta \omega$. In addition, this sampled noise is formally periodic, so $\Delta \omega ^{-1}$ must be much larger than the simulation time. 

\subsection{Trajectory Generation and Equilibrium Initialization}

To ensure the initial state at the simulation start time \(t=0\) corresponds to the correlated thermal equilibrium \(\rho_{SB}^\beta\), we must prepare corresponding equilibrium realization of trajectory \((x^{(j)}(t), p^{(j)}(t))\), which is correlated with a given noise realization \(\xi^{(j)}(t)\). We can do it in two ways.

The first approach relies on the ability of the system to equilibrate from an arbitrary initial state. So, the particle's trajectory \((x^{(j)}(t), p^{(j)}(t))\) is obtained by numerically integrating the classical GLE (derived from Eq.~\eqref{eq:gle_trajectory} for a free particle or in a potential \(V(x)\)):
\begin{equation}
    m \ddot{x}^{(j)}(t) = -V'(x^{(j)}(t)) - \int_{t_{start}}^{t} d\tau \, M(t-\tau) \dot{x}^{(j)}(\tau) + \xi^{(j)}(t).
    \label{eq:numerical_gle}
\end{equation}

\begin{enumerate}
    \item Begin simulations at a sufficiently distant past time \(t_{\text{start}} = -T_{\text{eq}}\), with \(T_{\text{eq}}\) chosen to be several times longer than the bath's memory time \(\tau_M\) and the system's relaxation time \(\gamma^{-1}\).
    \item  Initialize the particle with any convenient coordinates and momentum. Since the subsequent equilibration period $T_{eq}$ is chosen to be much longer than all relaxation timescales of the system, the particle's memory of its initial condition is completely erased before $t=0$, and the precise choice has no effect on the resulting statistics. For concreteness, we use the computationally simplest choice \(x(-T_{\text{eq}})=0, p(-T_{\text{eq}})=0\) throughout, though any other value yields identical results after equilibration. For the translationally invariant free particle, all initial positions are equivalent; we therefore choose \(x(-T_{\text{eq}})=0\) for simplicity.
    \item Evolve the system under Eq.~\eqref{eq:numerical_gle} from \(t_{\text{start}}\) to \(t=0\). This equilibration period allows the system-bath correlations to fully develop. The state at \(t=0\) is then statistically indistinguishable from the true thermal equilibrium state \(\rho_{SB}^\beta\).
    \item Store the particle's phase-space coordinates \((x^{(j)}(0^-), p^{(j)}(0^-))\) at the end of this equilibration run.
\end{enumerate}
This initialization procedure implicitly samples from the correct Wigner distribution \(W^{SB}(r, p, \{ r_i, p_i \}, -T_{eq})\) corresponding to \(\rho_{SB}^\beta\), circumventing the need for its explicit construction.  However, it essentially relies on the ability to equilibrate from an arbitrary initial state. It does not work for a free particle with super-Ohmic spectral density at $\alpha > 2$, where final momentum distribution differs from equilibrium. For such cases we can use another way.

The second approach is to explicitly generate an equilibrium trajectory. For full equilibrium we can write GLE (consider with $V(x) = m \omega_0^2 x^2/2$) in frequency representation, that gives expression for equilibrium trajectory through response function  $ \chi(\omega) = [m(\omega_0^2 - \omega^2 - i \omega \tilde \gamma (\omega))]^{-1}$ \cite{breuer}.

\begin{equation}
    x(\omega) = \chi(\omega) \xi(\omega)
\end{equation}

Thus, we generate a trajectory on the interval $(-T_{eq}, 0)$ by analogy with noise

\begin{equation}
    x(t_n) = \sqrt{\frac{\hbar \Delta \omega}{2\pi}} \sum_{k=-N}^{N} \sqrt{N(\omega_k)} \,  \chi(\omega_k) z_k \, e^{-i \omega_k t_n}.
    \label{eq:trajectory_eq}
\end{equation}

with the same realization of auxiliary noise $z$.

Both approaches have their advantages and disadvantages. The first way provides greater autonomy of our method, which may be useful in the long term. The second approach is applicable when the system does not thermalize or thermalizes too slowly.

Note also that we are not obliged to have a true equilibrium state throughout all of phase space. It is sufficient that the simulated state coincide with the equilibrium distribution only on the effective support of the preparation function.
For case of free QBM we can use translational invariance and shift each trajectory by any constant value to provide greater efficiency of equilibrium simulation.

\subsection{Monte Carlo Sampling of Preparations and Weighted Averages}
The action of a preparation operator at \(t=0\), described by the preparation function \(\lambda(r_i, p_i | \bar{r}, \bar{p})\) in the Wigner representation, is implemented as a reweighting and resampling step on the ensemble of equilibrated trajectories.
\begin{enumerate}
    \item For each trajectory \(j\), the coordinates \((\bar{r}^{(j)}, \bar{p}^{(j)}) = (x^{(j)}(0^-), p^{(j)}(0^-))\) serve as the input to the preparation function.
    \item A new set of post-preparation coordinates \((r_i^{(j)}, p_i^{(j)})\) is sampled. In this work, we employ a simple rejection sampling method: candidate points \((r, p)\) are drawn uniformly from a region encompassing the essential support of the target distribution (e.g., within several standard deviations of its mean), and accepted with a probability proportional to \(|\lambda(r, p | \bar{r}^{(j)}, \bar{p}^{(j)})|\).
    \item Each trajectory is assigned a complex weight
    \begin{equation}
        w^{(j)} = \lambda(r_0^{(j)}, p_0^{(j)} | \bar{r}^{(j)}, \bar{p}^{(j)}).
        \label{eq:trajectory_weight}
    \end{equation}
    \item The trajectory is then propagated forward from the new initial condition \((r_0^{(j)}, p_0^{(j)})\) at \(t=0^+\) under Eq.~\eqref{eq:numerical_gle}, generating its future evolution.
\end{enumerate}
For a sequence of interventions at times \(\{t_k\}\), this reweighting/resampling procedure is applied at each intervention time using the corresponding preparation function.
The expectation value of a particle observable \(\hat{O}\) with Weyl symbol \(O_W(x, p)\) at any time \(t > 0\) is computed via the weighted Monte Carlo average over the ensemble of \(N\) trajectories:
\begin{equation}
    \langle \hat{O}(t) \rangle = \frac{ \sum_{j=1}^{N} w^{(j)} \, O_W\left( x^{(j)}(t), p^{(j)}(t) \right) }{ \sum_{j=1}^{N} w^{(j)} }.
    \label{eq:final_observable}
\end{equation}
The statistical error scales as \(1/\sqrt{N}\), as is typical for Monte Carlo methods. This numerical framework provides a direct and computationally tractable path for simulating the exact non-Markovian quantum dynamics of Brownian motion for arbitrary temperatures and state preparations.

\section{Results and Comparison with Established Methods}
We now validate our stochastic numerical method and use it to investigate phenomena that are intractable for standard high-temperature approximations. First, we benchmark the method against exact analytical results for Gaussian preparations. We then employ it to study the decoherence of a non-classical state (a Schr\"odinger cat), comparing the exact quantum dynamics with the predictions of the common high-temperature master equation. Finally, we analyze the distinct thermalization behavior at low temperatures.

\subsection{Gaussian State Preparation}
As a first test, we consider a free QBM at $T=0$ with Gaussian preparation that localizes the particle's position which is centered at $\langle x \rangle = 0$ with initial spatial variance $\sigma_0^2$  \cite{Grabert1988}. The uncertainty principle mandates a corresponding increase in momentum variance, reflected in the Wigner representation:
\begin{equation}
    \lambda(r_0, \bar{r}, p_0, \bar{p}) =  \frac{\delta(r_0 - \bar{r})}{2(2\pi)^2} \exp\left(-\frac{r_0^2}{2\sigma_0^2} - \frac{2 \sigma_0^2}{\hbar^2} (p_0 - \bar{p})^2\right).
\end{equation}
For this preparation within the quadratic Caldeira-Leggett model, the dispersions of the coordinate and momentum have an exact analytical expression \cite{Grabert1988}:
\begin{gather}
    \sigma^2(t) = \langle x^2(t) \rangle = \sigma_0^2 + d^2(t) + \frac{A^2(t)}{\sigma_0^2}
    \label{eq:analytical_sigma}
\end{gather}
\begin{gather}
    \langle p^2(t) \rangle = \langle p^2 \rangle_{eq} + \frac{m^2 \dot A^2(t)}{\sigma_0^2} 
    \label{eq:analytical_p2}
\end{gather}
where $d^2(t) = \langle (x(t)-x(0))^2 \rangle$ is the mean squared displacement, $A(t) = (1/2i)\langle [x(t), x(0)] \rangle$ is the commutator response function and $\langle p^2 \rangle_{eq}$ is the equilibrium value of momentum dispersion.

\begin{figure*}[htbp]
\centering
\includegraphics[width=1.00\textwidth]{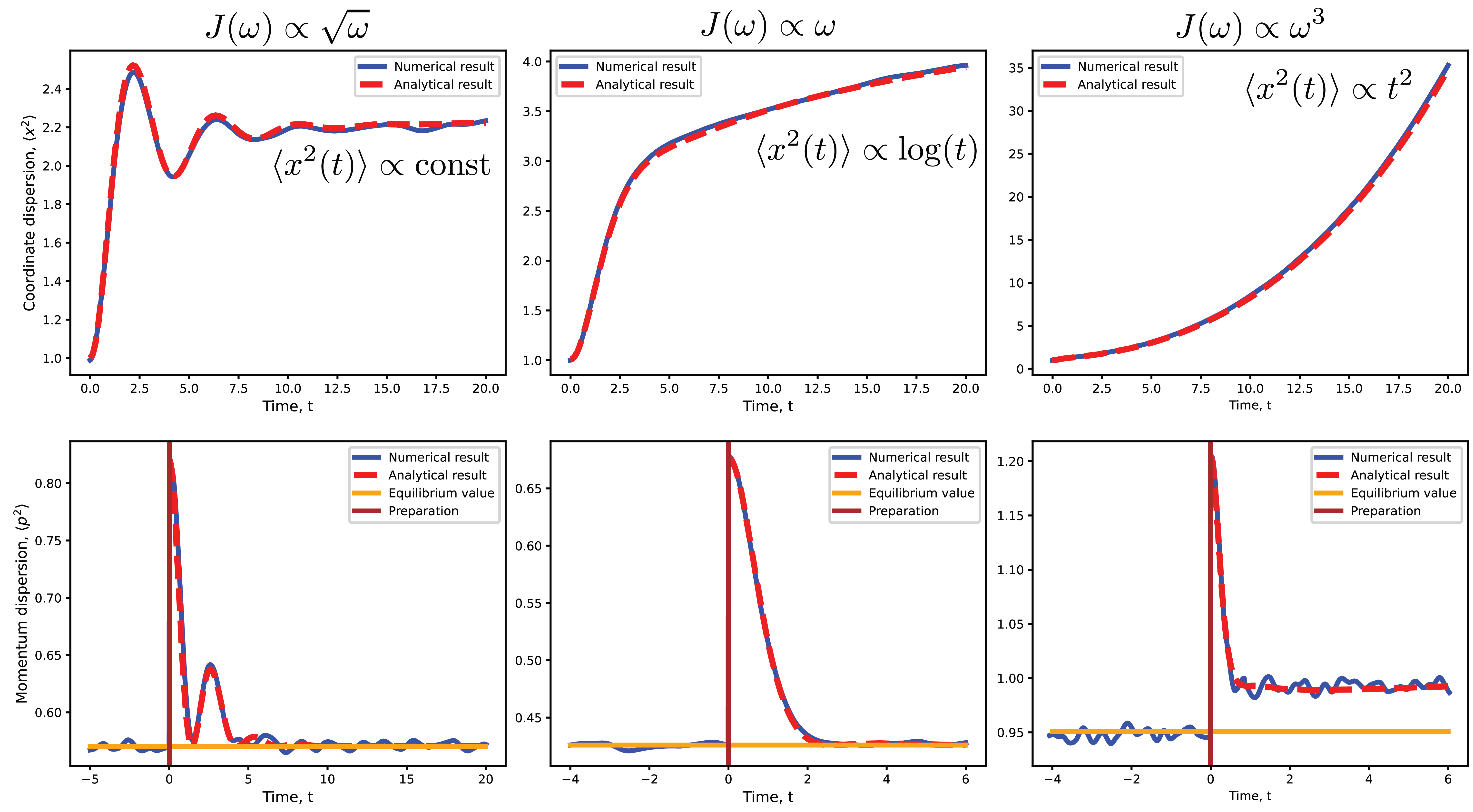}
\caption{Time evolution of the coordinate dispersion $\langle x^2(t) \rangle$ (top row) and momentum dispersion $\langle p^2(t) \rangle$ (bottom row) for a Gaussian preparation at $T=0$ with sub-Ohmic ($J(\omega) = g \sqrt{\omega} e^{-\varepsilon \omega}$), Ohmic ($J(\omega) = g \omega e^{-\varepsilon \omega}$) and super-Ohmic ($J(\omega) = g \omega^3 e^{-\varepsilon \omega}$) spectral densities. Blue line: stochastic Monte Carlo simulation. Red line: analytical result from Eq.\eqref{eq:analytical_sigma}, \eqref{eq:analytical_p2}. The agreement validates the numerical method. Momentum dispersion is shown both before and after preparation(brown line). For $\alpha = 1/2$ and $\alpha = 1$ relaxation to equilibrium value (orange line) is observed, in contrast to super-Ohmic density with $\alpha = 3 > 2$.  Parameters: $\sigma_0=1$, $g=1$, $\varepsilon=0.5$, $m=\hbar=1$ with $\Delta\omega = \frac{1}{600}$ and $\omega_m = 10 = 5\varepsilon^{-1}$.}
\label{fig:gaussian_dispersion}
\end{figure*}

Figure \ref{fig:gaussian_dispersion} shows the time evolution of the coordinate dispersion $\sigma^2(t)$ at zero temperature, computed using our stochastic method compared to the analytical formula Eq.~\eqref{eq:analytical_sigma}. The parameters are $\sigma_0=1$, $g = 1$, $\varepsilon=0.5$, and $m=\hbar=1$. The good agreement validates the core numerical implementation: our method of sampling noise from the zero-point spectral density and weighting trajectories via the Wigner preparation function accurately reproduces exact quantum results.

\subsection{Oscillator in the Ohmic bath at equilibrium}

\begin{figure}[htbp]
\centering
\includegraphics[width=0.48\textwidth]{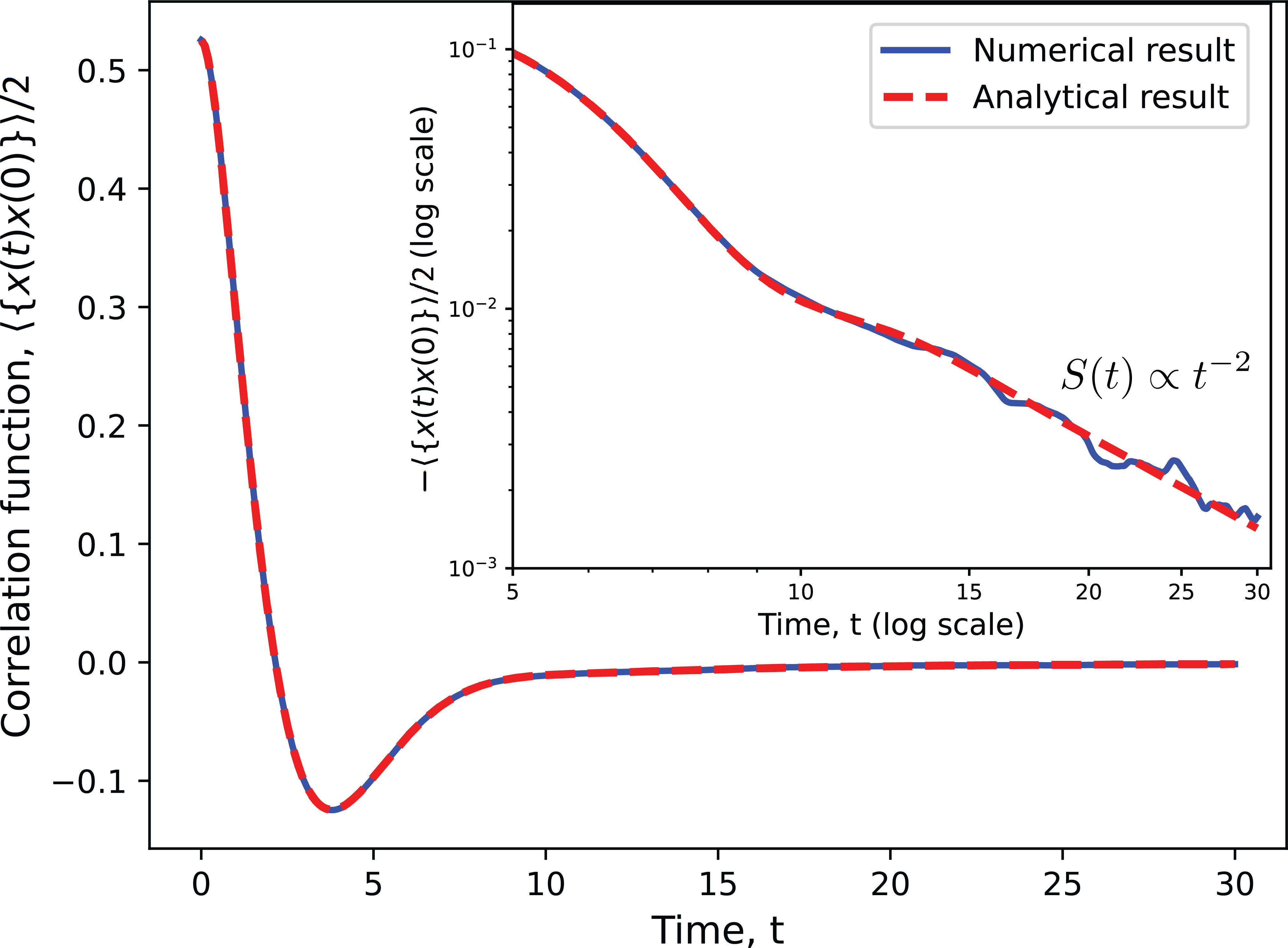}
\caption{Correlation function $S(t) = \langle \{ x(t) x(0)\} \rangle / 2$ for oscillator at equilibrium in Ohmic bath ($J(\omega) = m \gamma \omega e^{-\varepsilon \omega}$) at zero temperature $T = 0$.  Blue line: stochastic Monte Carlo simulation. Red line: analytical result \cite{breuer, Grabert1988}. The agreement validates the numerical method. The inner plot shows the numerical and analytical results (absolute values) on a log-log scale. A negative algebraic asymptotic behavior—essentially quantum behavior—is observed, which is reproduced by our method. Parameters:  $\gamma = 1$, $\Omega  = 1/\sqrt{2}$, $\varepsilon = 0.1$, $m=\hbar=1$ with $\Delta \omega = \frac{1}{200}$ and $\omega_m = 50 = 5\varepsilon^{-1}$.} 
\label{fig:oscillator}
\end{figure}

As a second test, we consider the harmonic oscillator with frequency $\Omega$ at $T = 0$ in the Ohmic bath $J(\omega) = \gamma m \omega e^{-\varepsilon \omega}$. An interesting quantity is an equilibrium correlation function $S(t) = \frac{1}{2} \langle \{x(t)x(0) \} \rangle$ \cite{breuer, Grabert1988}. While in the classical case it decays exponentially, in the essentially quantum regime at zero temperature this decay is algebraic \cite{gardiner2004quantum}, which is reproduced by our method, as can be seen in Figure \ref{fig:oscillator}. The parameters are $\gamma=1$, $\varepsilon=0.1$, $\omega_0 = 1/\sqrt{2}$ and $m=\hbar=1$.

\subsection{Decoherence Dynamics: Schr\"odinger Cat State}
To demonstrate the capability of our method for non-classical states and to highlight the limitations of high-temperature approximations, we study the decoherence of a Schr\"odinger cat state for free QBM. The state is a superposition of two spatially separated Gaussian wave packets:
\begin{equation}
\begin{gathered}
    |Cat\rangle = \frac{1}{\sqrt{\mathcal{N}}}( |+\rangle + |-\rangle ), \\     
    \langle x|\pm\rangle = \frac{1}{(2\pi\sigma^2)^{1/4}} \exp\left(-\frac{(x \pm x_0)^2}{4\sigma^2}\right),
\end{gathered}
\end{equation}

where $\mathcal{N} = 2(1 + \langle -|+\rangle)$. The corresponding preparation function in the Wigner representation factorizes: $\lambda(r, p| \bar{r}, \bar{p}) = \mathcal{Z}^{-1} W_{\text{Cat}}(r, p) W_{\text{Cat}}(\bar{r}, \bar{p})$,  where

\begin{multline}
    W_{Cat}(r,p) = \frac{\exp{-2\sigma^2 p^2}}{\pi \mathcal{N}} \bigg( \exp{-\frac{(r-x_0)^2}{2\sigma^2}} + \\ +\exp{-\frac{(r+x_0)^2}{2\sigma^2}}  + 2 \exp{-\frac{r^2}{2\sigma^2}} \cos(2 x_0 p)\bigg) 
\end{multline}

Note that $W_{Cat}$ can take negative values, which completely excludes the possibility of interpreting it as a probability. However, the offered method of sampling of initial conditions handles it. 

We compute the decay of the coherence element, represented by the observable $\hat{O} = |+\rangle\langle -| + |-\rangle\langle +|$, whose Weyl symbol is
\begin{equation}
    O_W(r, p) = 4 \exp\left( -\frac{r^2}{2\sigma^2} - 2\sigma^2 p^2 \right) \cos(2 x_0 p).
    \label{eq:cat_observable}
\end{equation}

\begin{figure}[htbp]
\centering
\includegraphics[width=0.51\textwidth]{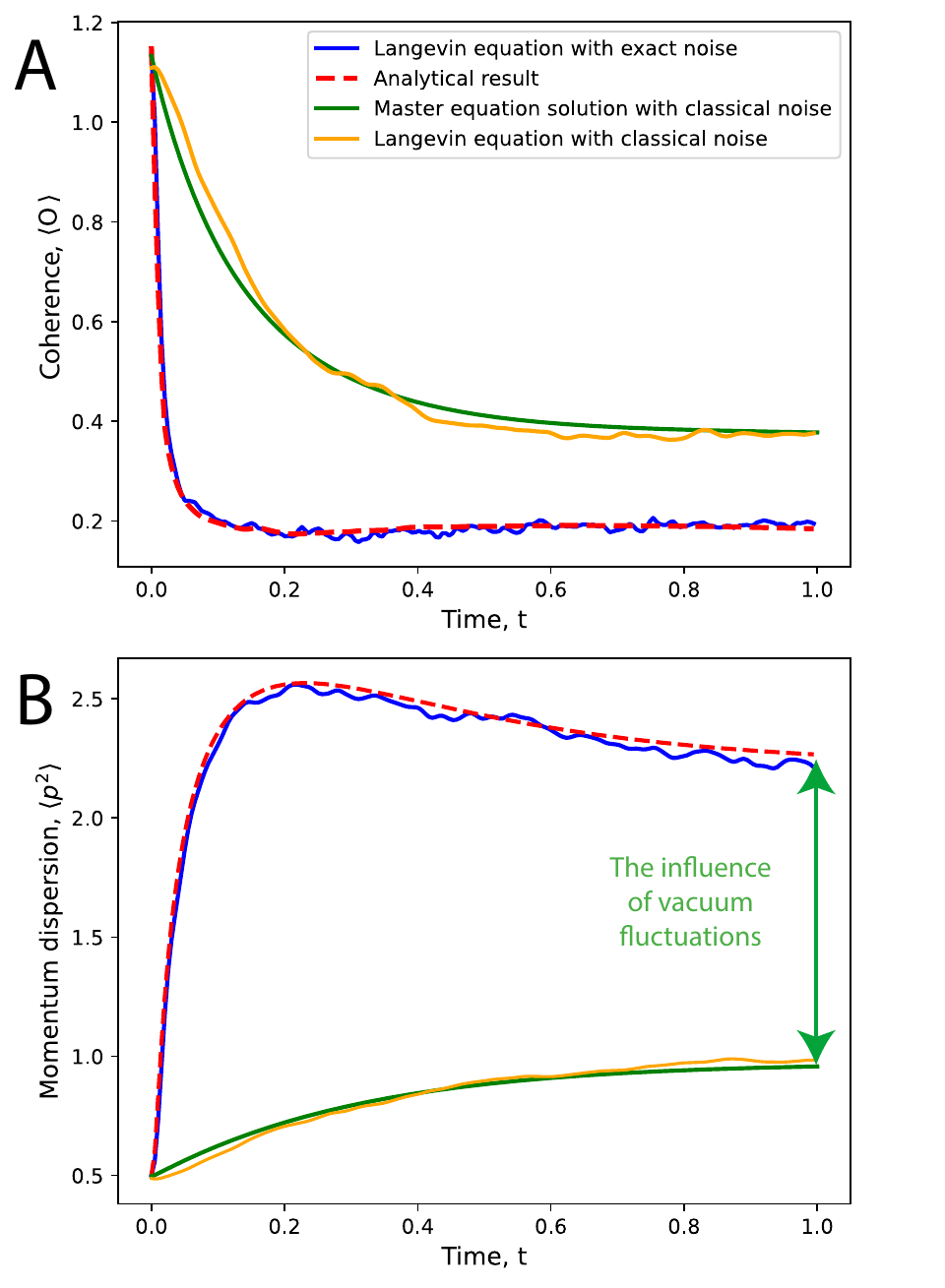}
\caption{Evolution of a Schr\"odinger cat state. \textbf{A}. The coherence $\langle O \rangle$ (Eq.~\eqref{eq:cat_observable}) decays fastest for the exact quantum noise (blue  - numerical result with exact noise and red - analytical result (Appendix \ref{cat_app})). White-noise process (green -  the high-temperature master equation and orange - Langevin equation with classical noise) give slower decoherence, failing to capture the enhanced quantum fluctuations at low temperature. \textbf{B}. Momentum dispersion $\langle p^2(t) \rangle$ relaxation for Schr\"odinger cat state. The exact quantum result (blue and red lines, stochastic and analytical) shows rapid initial growth on a timescale $\sim \varepsilon$, absent in the high-temperature white-noise model (orange and green). This demonstrates the quantum bath's role in fast early-time dynamics. Parameters: $\gamma=\pi/2$, $\varepsilon=0.01$, $T=1$, $m=\hbar=1$, $x_0=1$, $\sigma = 1/2$. Frequency grid has $\omega_m = 200 = 2 \varepsilon^{-1}$ and $\Delta\omega = \frac{1}{600}$.} 
\label{fig:cat_decoherence}
\end{figure}

We compare three distinct dynamics in Fig. \ref{fig:cat_decoherence}:
\begin{enumerate}
    \item {Exact Quantum Dynamics (Our method):} The colored noise correlation uses the full quantum FDT, Eq.~\eqref{eq:noise_target}.
    \item {High-Temperature White Noise:} The noise correlation uses the classical FDT, $\langle \xi(t)\xi(0)\rangle \propto 2m\gamma  T \delta(t)$, corresponding to the standard Markovian limit.
    \item {High-Temperature Master Equation (ME):} Dynamics from the common Caldeira-Leggett master equation \cite{Caldeira1983, breuer}:
    \begin{equation}
        \dot{\rho} = -\frac{i}{\hbar}[H_0, \rho] - \frac{i\gamma}{2\hbar}[x, \{p, \rho\}] - \frac{m\gamma T}{\hbar^2}[x, [x, \rho]].
        \label{eq:highT_ME}
    \end{equation}
\end{enumerate}
The parameters are $\gamma=\pi/2$, $\varepsilon=0.01$, $T=1$ and $m=\hbar=1$. Crucially, for these parameters, $T \sim 1$ is not much greater than the effective cutoff energy $\hbar/\varepsilon \approx 100$, violating the condition ($T \gg \hbar/\varepsilon$) required for the Markovian/high-temperature approximation to be valid \cite{Ferialdi}.

Figure \ref{fig:cat_decoherence} reveals significant differences. The high-temperature ME and white-noise stochastic process predict nearly identical, relatively slow decoherence. In stark contrast, the exact quantum dynamics exhibits much faster initial decoherence. This is a direct manifestation of zero-point and low-temperature fluctuations in the bath, which are neglected in the classical FDT. Our method, which incorporates the full quantum noise spectrum, is essential to capture this correct physics.

\subsection{Problems with high-temperature Markovian Master-Equation with white noise}

In this section, we discuss why the Lindblad equation is not applicable to QBM.

The classical Caldeira-Leggett model admits a Markovian limit with white noise, requiring only one condition:  $\varepsilon \ll \min \{\gamma^{-1}, \tau_S \}$. It follows $\langle \xi(t) \xi(s) \rangle \propto \delta(t-s)$. In quantum case, with finite $\hbar$,  for (\ref{eq:FDT}) to become a delta function, an additional condition is required: $T \gg \hbar / \varepsilon$. If we want to consider non-negligible damping, Markovianity requires a very small $\varepsilon$. Simultaneously, if the system is genuinely quantum ($\hbar \sim 1$)  temperature must be even more than already big $1/\varepsilon$. So, in this case, this ME correctly describes QBM with effectively infinite temperature:  $T \gg \varepsilon^{-1} \gg \gamma$. 

If we work with finite temperatures, we must take into account colored noise for the correct description of essentially quantum and strongly damped systems. It should be noted that white noise is still applicable if we are working with weak damping, as in quantum optics, where the characteristic relaxation times ($\gamma^{-1}$) are much larger than $\varepsilon$ and $\tau_S$. 

The nature of this problem is rooted in vacuum fluctuations. Classically, at $T=0$ all  oscillators have zero squared coordinates and momenta. In the quantum case, however, they are constrained by the uncertainty principle. So even when thermal fluctuations are negligible, vacuum noise remains. Furthermore, the oscillator dispersions have a different frequency dependence, which leads to a different, essentially non-Markovian noise statistic (\ref{eq:FDT}). In particular, the amplitude of the vacuum noise (unlike thermal noise) increases with spectral width. Finally, to reach the Markovian limit, we must increase this width, which in turn increases vacuum noise. Consequently, we must choose the temperature large enough so that white noise becomes much stronger than the vacuum noise. 

If we consider only white noise for finite temperature, we will encounter two problems. Firstly, vacuum noise has a significant influence on the dynamics, as can be seen in Fig. \ref{fig:cat_decoherence}. Secondly, white noise is essentially unphysical in the quantum case. 

Let us illustrate this with an oscillator at finite temperature. If we consider equilibrium values of coordinate and momentum dispersions, uncertainty principle requires the condition $T \geq \hbar \omega_0 / 2$, which also limits applicability of white noise to low temperatures in principle.

\subsection{Low-Temperature Thermalization and the Role of the Cutoff}
The failure of the high-temperature approximation is further elucidated by examining momentum thermalization. Fig. \ref{fig:cat_decoherence} (B) shows the time evolution of $\langle p^2(t) \rangle$ starting from an initial value of zero. The exact quantum result (from both analytical integration \cite{breuer} and our stochastic method) shows a rapid initial increase on a timescale $t \sim \varepsilon$, followed by a slower approach to the final equilibrium value. The high-temperature (white-noise) prediction lacks this initial rapid rise entirely, as it misses the dominant contribution from high-frequency bath modes ($\omega \sim \varepsilon^{-1}$).

This behavior can be understood analytically from the low-temperature limit of the momentum variance \cite{breuer}:
\begin{equation}
\label{eq_momentum}
    \langle p^2(t) \rangle \approx \frac{m \gamma \hbar}{\pi} \int_0^\infty d\omega \, \omega e^{-\varepsilon \omega} \frac{|e^{i\omega t} - e^{-\gamma t}|^2}{\omega^2 + \gamma^2}.
\end{equation}
For short times $t \ll \varepsilon$, the integral yields $\langle p^2(t) \rangle - \langle p^2(0) \rangle \propto (m \gamma \hbar / \pi) (t^2/\varepsilon^2)$. For intermediate times $\varepsilon \ll t \ll \gamma^{-1}$, the growth is logarithmic: $\propto (2m \gamma \hbar / \pi) (1-\gamma t) \ln(t/\varepsilon)$. Thus, the bath cutoff $\varepsilon^{-1}$ sets the timescale for the initial, rapid thermalization driven by vacuum fluctuations—a genuinely quantum effect that ensures fast decoherence even at $T=0$.

 \section{Extension to Non-Quadratic Potentials: A Controlled Approximation}
The exact quantum-classical correspondence demonstrated in the previous sections holds strictly for Hamiltonians that are quadratic in both the particle and bath coordinates. For an arbitrary external potential \( V(x) \), this exact correspondence is broken. Remarkably, however, the intrinsic structure of the Caldeira-Leggett model provides a natural mechanism that suppresses quantum coherence, rendering the dynamics increasingly classical. This suggests that the stochastic approach may be extendable as a controlled approximation.

\subsection{A Natural Small Parameter: The Equilibrium Coherence Length}
The key lies in analyzing the equilibrium reduced density matrix of the particle, \(\rho_S^{\text{eq}}(x, x') = \rho_S^{\text{eq}}(r, q)\), derived from the full thermal state \(\rho_{SB}^\beta\). For the Caldeira-Leggett model, this density matrix develops an exponential decay in its off-diagonal coordinate \(q = x - x'\) \cite{Weiss_2012}:
\begin{equation}
    \rho_S^{\text{eq}}(r, q) \propto \exp\!\left(-\frac{1}{2\hbar^2} \langle p^2 \rangle_{\text{eq}} \, q^2 \right),
    \label{eq:quasidiagonal_form}
\end{equation}
where \(\langle p^2 \rangle_{\text{eq}}\) is the equilibrium momentum variance of the particle, given by (\ref{eq_momentum}) in the limit of a wide bath spectrum. This defines a coherence length \(\lambda\):
\begin{equation}
    \lambda \equiv \frac{\hbar}{\sqrt{\langle p^2 \rangle_{\text{eq}}}}.
    \label{eq:coherence_length}
\end{equation}
The physical significance of \(\lambda\) is clear from Eq.~\eqref{eq:quasidiagonal_form}: the reduced density matrix becomes strongly peaked along the diagonal \(q=0\) (i.e., \(x \approx x'\)) when \(\lambda\) is small. The off-diagonal elements, which encode quantum coherence, are exponentially suppressed for \(|q| \gg \lambda\).

Crucially, for an ohmic bath with a high-frequency cutoff \(\Lambda \sim \varepsilon^{-1}\), the momentum variance diverges in the wide-band limit:
\begin{equation}
    \langle p^2 \rangle_{\text{eq}} \propto \int_0^{\Lambda} d\omega \, \frac{J(\omega)}{\omega} \coth\left(\frac{\hbar \omega}{2 T}\right) \xrightarrow[\Lambda \to \infty]{} \infty.
\end{equation}
Consequently, the coherence length vanishes:
\begin{equation}
    \lambda \xrightarrow[\Lambda \to \infty]{} 0.
    \label{eq:lambda_vanishes}
\end{equation}
This occurs even at zero temperature due to coupling to vacuum fluctuations, unlike the thermal coherence length \(\lambda_{\text{th}} \sim \hbar / \sqrt{m T}\), which remains finite as \(T \to 0\). Thus, \(\lambda\) emerges as a natural, bath-controlled small parameter that quantifies the degree of "classicality" induced in the particle by its interaction with a broad-spectrum environment.

\subsection{Possibility of the Stochastic Approximation}
We expect that the stochastic approach derived for quadratic potentials can be extended to a smooth, non-quadratic potential \(V(x)\) when the system operates in the quasi-classical regime. This regime is defined by the condition that the potential varies slowly over the scale of the quantum coherence length \(\lambda\).

To assess the possibility of applying this method to a non-quadratic external potential, consider effective action in the influence functional approach \cite{Feynman1963, Grabert1988}. Non-quadratic terms are located in $V(r+q/2) - V(r-q/2)$ (\ref{path_int}), where $r$ and $q$ are "center of mass" variables, introduced above. Assuming potential $V$ is smooth, expand this expression
\begin{equation}
    V'(r)q + \frac{V'''(r)}{24} q^3 + ...
\end{equation}

If we have only the first term, exponent with effective action  can be integrated explicitly, which leads to purely classical trajectories. Therefore we should be able to neglect the terms with $q^3$ and beyond. 

Because of noise, path integral is dominated by $q(s)$ close to zero with characteristic scale $\lambda$. This leads to the following simplest estimate: $V'(r) \lambda \gg V'''(r) \lambda^3$ for all $r$ (or $r$ in characteristic area of motion of the Brownian particle). Thus we can introduce  the characteristic classical length scale of the potential \(L = \min |V'(x)/V'''(x)|^{1/2}\).

Our conjecture that the error incurred by approximating the exact quantum dynamics with the classical stochastic process is controlled by the dimensionless ratio
\begin{equation}
    \epsilon \equiv \frac{\lambda}{L} \ll 1.
    \label{eq:small_parameter}
\end{equation}
When \(\epsilon \ll 1\), the particle's density matrix remains nearly diagonal throughout its evolution, as any generated coherences are rapidly suppressed by the bath on the scale \(\lambda\). In this regime, the dominant contributions to path integrals or master equations come from nearly diagonal histories, which are accurately captured by the ensemble of classical trajectories governed by the GLE with quantum noise.

Therefore, for potentials satisfying Eq.~\eqref{eq:small_parameter}, we expect the stochastic method to provide an excellent approximation to the true quantum dynamics, with deviations of order \(\mathcal{O}(\epsilon)\). This generalizes the utility of our approach beyond exactly solvable models, providing a powerful and practical tool for studying dissipative quantum dynamics in complex potentials, from chemical reaction profiles to disordered systems.

\subsection{Outlook and Future Development}
This analysis provides a clear pathway for future work. The next step is to develop a systematic perturbation theory in the small parameter \(\epsilon = \lambda/L\). This would allow for the calculation of leading quantum corrections to the classical stochastic dynamics, potentially through the inclusion of non-local (in time) correction terms to the stochastic weights or via a modified Fokker-Planck equation.  If successful, this framework would bridge the gap between the exact correspondence established for quadratic systems and the approximate but highly efficient stochastic simulation available for a much broader class of problems.

\section{\label{sec:concl}Conclusions}

The central challenge in simulating quantum Brownian motion (QBM) is the exact treatment of environmental influence, which makes direct quantum mechanical calculations intractable due to the infinite-dimensional nature of the bath's Hilbert space.  Numerically exact methods such as the hierarchical equations of motion (HEOM) and the stochastic Schrödinger equation address this challenge, but face increasing computational demands at low temperatures or for unconfined motion with large effective state-space dimension. While the path-integral framework provides an exact solution for the quadratic Caldeira-Leggett model, the resulting expressions are cumbersome and resist extension to complex preparation protocols.

In this work, we have introduced a profound reinterpretation and a powerful numerical method that overcomes these barriers. We have demonstrated that the exact quantum dynamics of a particle in the Caldeira-Leggett model can be mapped, at any temperature, onto a classical, non-Markovian stochastic process in phase space. For a particle with a quadratic Hamiltonian, this correspondence is proven exact, starting from the physically correct correlated thermal equilibrium state and accommodating arbitrary state preparations and interventions via the Wigner representation of preparation functions.

The primary value of this mapping is its profound simplification: it reduces the formidable problem of simulating an open quantum system to the tractable task of generating and averaging classical stochastic trajectories. We have developed and validated a corresponding Monte Carlo numerical technique. The method correctly reproduces established analytical results and, crucially, captures the fast decoherence and thermalization dynamics driven by vacuum and low-temperature fluctuations—effects that are missed by common high-temperature master equations.

Furthermore, we have identified a natural small parameter that governs the extension of this framework: the equilibrium coherence length \(\lambda = \hbar / \sqrt{\langle p^2 \rangle_{\text{eq}}}\), which shrinks with increasing bath spectral width.  We conjecture that for smooth, non-quadratic potentials where the classical scale \(L\) satisfies \(\lambda/L \ll 1\), our stochastic approach provides a controlled approximation, with error of order \(\mathcal{O}(\lambda/L)\).  If confirmed, this opens a clear pathway for future development, including systematic perturbation theory or hybrid numerical schemes that treat confined quantum dynamics exactly within the coherence width \(\lambda\) while sampling the larger-scale classical motion stochastically.

In summary, we have established a versatile, all-temperature computational framework for QBM  with at most quadratic potential. By translating non-Markovian quantum dynamics into a classical stochastic language, we provide a conceptually simpler and numerically efficient alternative to more complex contemporary methods, offering a unified tool for studying dissipative quantum dynamics from the classical high-temperature regime down to the deeply quantum zero-temperature limit.

\subsection*{Acknowledgement}
This work was supported by Rosatom in the framework of the Roadmap for Quantum computing (Contract No. 868-1.3-15/15-2021 dated October 5). 
\newpage

\appendix
\section{Influence functional}
\label{pathint}
In this section, we show the equality of the GLE approach and the exact quantum solution using the influence functional \cite{Grabert1988}. We consider the propagator, which is explicitly given there, with the following changes.

Note that $\rho_\beta^{SB}$ is a stationary state of the full Hamiltonian, so we can identically rewrite it as
\begin{equation}
    U(0, -T)\rho_\beta^{SB} U (-T, 0).
\end{equation}

Thus, we can carry out the integration in imaginary time back to $-T$. The preparation remains at zero. This will lead to a discontinuity in the particle trajectories, but, due to the fact that the preparation does not affect the thermostat, the bath oscillator trajectories will remain continuous, and after taking a partial trace, they will also enter into effective action in the form of corresponding kernels, but with a history from $-T$ to $t$ instead of $(0,t)$. So this propagator

\begin{equation}
\begin{gathered}
\label{path_int}
J(q_f, r_f, t, q_i, r_i, \bar{r}, \bar{q}) = Z^{-1} \int \! \mathcal{D}q \, \mathcal{D}r \, \mathcal{D}\bar{x} \\ \exp \frac{i}{\hbar}\Bigg(i \int_{0}^{\hbar\beta} d\tau \Big[ \frac{m}{2} \dot{\bar{x}}^2 + V(\bar x) + \frac{1}{2} \int_{0}^{\hbar\beta} d\sigma \, k(\tau - \sigma) \bar{x}(\tau) \bar{x}(\sigma) \Big] \\ + \int_{0}^{\hbar\beta} d\tau \int_{-T}^{t} ds \, K^*(s+T - i\tau) \bar{x}(\tau)q(s)  +\\ \int_{-T}^{t} ds \Big[ m\dot{q}\dot r - V(r+\frac{q}{2}) + V(r-\frac{q}{2}) \\ - \int_{-T}^{s} du  M(s - u)q(s) \dot{r}(u) - r(-T) M(s+T)q(s) \Big] \Bigg)\\ \exp{- \frac{1}{2 \hbar^2} \int_{-T}^{t} ds   \int_{-T}^{t} du \, N(s - u)q(s)q(u)}
\end{gathered}
\end{equation}

where $q_f = q(t)$, $r_f = r(t)$, $q_i = q(+0)$, $r_i = r(+0)$, $\bar q = q(-0)$, $\bar r = r(-0)$, integration over $\mathcal{D} \bar x$ with $\bar x(\hbar \beta) = r(-T) + q(-T)/2$ and $\bar x(0) = r(-T) - q(-T)/2$; $k$ and $K^*$ is kernels, introduced in \cite{Grabert1988}. For our calculations it is enough that $K(t-i\tau) \xrightarrow[]{}0 $ at $t\xrightarrow[]{} \infty$. It means that dynamics at $t>0$ connected only with real-time dynamics at $t<0$ if $T \xrightarrow[]{} \infty$. In turns, dynamics at $t<0$, which is connected with imaginary time path, is trivial dynamics of the stationary equilibrium state.

Firstly, we introduce noise by Hubbard–Stratonovich transformation. 

\begin{equation}
\begin{gathered}
    \exp{- \frac{1}{2 \hbar^2} \int_{-T}^{t} ds   \int_{-T}^{t} du \, N(s - u)q(s)q(u)} = \int \mathcal{D} \xi \\ \exp{-\frac{1}{2} \int_{-T}^t \int_{-T}^t ds du \xi(s) N^{-1}(s-u) \xi(u) +  \frac{i}{\hbar} \int_{-T}^t q(s) \xi(s) ds}
\end{gathered}
\end{equation}

Secondly, we integrate by parts the following term, accounting the discontinuity at $t=0$. 

\begin{equation}
\begin{gathered}
    \int_{-T}^{t} ds  m\dot{q}\dot r = \bigg( \int_{-T}^0 + \int_0^t \bigg) ds  m\dot{q}\dot r = - \int_{-T}^t ds q(s) m \ddot r(s)\\+ m(q(t)\dot r(t) - q(+0) \dot r(+0) + q(-0) \dot r(-0) - q(-T) \dot r(-T)) = \\ = - \int_{-T}^t ds q(s) m \ddot r(s) + m(q_f \dot r_f - q_i \dot r_i + \bar q \dot {\bar r} - q_{-T} \dot r_{-T})
\end{gathered}
\end{equation}

Wigner transformation for $q_f, q_i, \bar q$ leads to boundary conditions $\delta(m\dot r_f - p_f) \delta(m\dot r_i - p_i) \delta(m\dot {\bar r} - \bar p)$. Since $V$ is at most quadratic, $V(r+q/2) - V(r-q/2) = q V'(r)$. After that we can integrate by $\mathcal{D} q$, that follows to $\delta(m\ddot r(s) + V'(r(s)) + \int_{-T}^s du M(s-u) \dot r(u) - \xi(s))$.

For $s<0$ we immediately assume only one saddle trajectory, corresponding $(\bar r, \bar p)$ at $t = -0$ and given realization $\xi(t)$. Thus, we obtain 

\begin{equation}
\begin{gathered}
J_W(p_f, r_f, t, p_i, r_i, \bar{r}, \bar{p}) = Z^{-1} \int \mathcal{D} \xi \\ \exp{-\frac{1}{2} \int_{-T}^t \int_{-T}^t ds du \xi(s) N^{-1}(s-u) \xi(u)} \\ \int \mathcal{D} r \delta(m\dot r_f - p_f) \delta(m\dot r_i - p_i) \delta(m\dot {\bar r} - \bar p) \\ \delta \bigg(m\ddot r(s) + V'(r(s)) + \int_{-T}^s du M(s-u) \dot r(u) - \xi(s) \bigg)
\end{gathered}
\end{equation}

We have obtained the analogous representation for classical GLE \cite{Hanggi1993}, which is also corresponding with our numerical approach. 

\section {Calculation of coherence observable for Schrodinger cat state}
\label{cat_app}
The initial state is pure and has the form $\psi_+ + \psi_-$. Thus, the Wigner function is not Gaussian for the full state. However, it can be represented as $\propto W^{++} + W^{-+} + W^{+-} + W^{--}$, where  each part is formally Gaussian. Because of the Gaussian evolution, these parts will also be Gaussian at arbitrary $t>0$. So, we can consider separately $W^{++}(x, p)$  and $W^{+-}(x, p)$ as independent normalized states (due to the symmetry) and obtain

\begin{equation}
    \langle O \rangle = \frac{\langle O \rangle^{++}}{1 + \langle + | - \rangle} + \frac{\langle O \rangle^{+-} \langle + | - \rangle}{1 + \langle + | - \rangle} 
\label{Cat}
\end{equation}

Gaussian states are expressed through first and second momenta

\begin{equation}
\begin{gathered}
    W(x,p) = \frac{1}{2\pi\sqrt{D}} \exp(-\frac{\tilde x^2 \sigma_{pp} - 2\tilde x\tilde p \sigma_{xp} + \tilde p^2 \sigma_{xx} }{2D})
\end{gathered}
\end{equation}

where $\tilde x = x-\langle x \rangle $, $\tilde p = p-\langle p \rangle $, $\sigma_{xx} = \langle \tilde x^2 \rangle$, $\sigma_{xp} = \langle \tilde x \tilde p \rangle$, $\sigma_{pp} = \langle \tilde p^2 \rangle$, $D = \sigma_{xx} \sigma_{pp} - \sigma_{xp}^2$.

Average values are obtained by direct integration

\begin{equation}
    \langle O \rangle^{++, +-} = \int dx dp \, O(x,p) W^{++, +-}(x,p),
\end{equation}

 where the moments obtained from the solution of the Heisenberg equations \cite{breuer} are substituted. Note that for $+-$ term $\langle x \rangle$ and $\langle p \rangle$ are imaginary.

\end{document}